\documentstyle[epsfig,twocolumn,seceq]{asr2k}

\title
{
Static Spin Correlation in LTT Phase of La$_{1.875}$Ba$_{0.075}$Sr$_{0.05}$CuO$_4$
}
\def\runtitle
{
Static Spin Correlation in LTT Phase of La$_{1.875}$Ba$_{0.075}$Sr$_{0.05}$CuO$_4$
}

\author
{
Masaki {\sc Fujita}, Hideto {\sc Goka}, Kazuyoshi {\sc Yamada}
}

\def\runauthor
{
Masaki {\sc Fujita}, Hideto {\sc Goka} and Kazuyoshi {\sc Yamada}
}

\inst
{
Institute for Chemical Research, Kyoto University, Gokasyo, Uji 611-0011, Kyoto, Japan
}
\abst
{
We have carried out elastic neutron scattering measurements on La$_{1.875}$Ba$_{0.075}$Sr$_{0.05}$CuO$_4$ single crystal ($T_c\approx$10K). Incommensurate elastic magnetic peaks were observed in the low-tamperature tetragonal phase with the propergation vector parallel/perpendicular to in-plane Cu-O bond direction. The magnetic peak intensity normalized by the sample volume is approximately six times larger than that of the orthorhombic La$_{1.88}$Sr$_{0.12}$CuO$_4$ at low temperature.
}

\kword
{
superconductivity, LTT structure, static spin correlation,  neutron scattering
}
\begin{document}
\sloppy
\maketitle
\section{Introduction}
It is well known that the superconductivity is anomalously suppressed near the hole concentration of 1/8 per Cu atom. (1/8 problem)$^{1,2)}$ The mechanism of suppression is studied actively in connection with that of high-$T_c$ superconductivity.

Recent neutron scattering measurements reveal evidences of static order of both spin and charge in La$_{1.48}$Nd$_{0.4}$Sr$_{0.12}$CuO$_{4}$.~\cite{Tranquada95} Then the stripe structure of holes was proposed for understanding the 1/8 problem. Subsequently, the magnetic ordering was confirmed in the orthorhombic La$_{1.88}$Sr$_{0.12}$CuO$_{4}$$^{4,5)}$ and La$_{2}$CuO$_{4+y}$~\cite{YoungLee99} phases. The observed incommensurate (IC) peak positions in the Nd-doped system form a square,~\cite{Tranquada96} while in the latter two Nd-free systems the positions shift to a rectangular arrangement.$^{5,6,8)}$ Furthermore, the superlattice reflection associated with the charge order has not been confirmed in the low-temperature orthorhombic(LTO) phase. It is therefore important to clarify the relationship among the stripe order, the crystal structure and the superconductivity.
To address this issue, we performed neutron scattering study on  La$_{1.875}$Ba$_{0.075}$Sr$_{0.05}$CuO$_4$ single crystal. As shown later, present crystal shows the transition to low-temperature tetragonal(LTT) phase at low temperature. This study is therefore the first neutron scattering study on single crystal for investigating the physical properties in LTT phase without rare-earth moment.
\section{Experimental}
Single crystal of La$_{1.875}$Ba$_{0.075}$Sr$_{0.05}$CuO$_4$ was grown by a travelling-solvent floating-zone method. The crystal growth was done under the almost same conditions for growing La$_{2-x}$Sr$_{x}$CuO$_4$ single crystal~\cite{Hosoya94}. For neutron scattering measurements crystal rod with diameter of 6mm is cut into 25 mm in length. The crystal was annealed in oxygen gas flow at 900 $^{\circ}${\it C} for 50 hours and subsequently at 500 $^{\circ}${\it C} for 50 hours, followed by gradual cooling in the furnace. Then the field-shielding effect was measured using a SQUID magnetometer under magnetic field of 10 Oe. From the result shown in Fig.1(a), the onset temperature for superconductivity $T_c$(onset) is determined to be 10 K, consistent with the previous result for powder sample.~\cite{Maeno91}

Neutron scattering measurements were carried out at the KSD double-axis spectrometer and the HER triple-axis spectrometer installed in the JRR-3M Guide Hall at the JAERI in Tokai, Japan. The incident neutron beams with the fixed energy of 34.9 meV and 5 meV are used at the KSD and the HER spectrometers, respectively. The horizontal divergence was set to be 12' for incident beam and to be 15' and 40' for scattered one at KSD. The horizontal collimation sequences of 32'-100'-80'-80' were used at HER. The sample was mounted so that the ({\it h},{\it k},0) plane is in the scattering plane. All crystallographic indexes in this paper are described in the tetragonal ({\it I}4/{\it mmm}) notation. 
\section{Results}
\subsection{Structural phase transition}

In order to confirm the phase transition from LTO to LTT, we investigated the profile of fundamental Bragg peak. Fig.1 (b) shows the elastic profiles through (2,2,0) at 12K and 50K. The single (2,2,0) Bragg peak at 12 K in LTT phase splits into four peaks at 50 K in LTO phase because of the domains. In this sample, we confirmed the existence of four equally distributed domains in LTO phase. For the determination of transition temperature ($T_{d2}$), temperature dependence of peak intensities at (2,2,0) and (2.014,2,0) were measured. The results are depicted in Fig.1 (c). The evaluated $T_{d2}$ of 37 K is identical with the result for powder sample.~\cite{Maeno91} 

\subsection{Static IC spin correlation}
We observed the elastic magnetic peaks around (0.5,0.5,0) at low temperature. Fig.2 shows the elastic spectrum scanned along (0.5,{\it k},0). In the figure, open and closed symbols are data at 4.5 K and 45 K, respectively. For the quantitative analysis, two Lorentzian functions, 
\begin{displaymath}
   \frac{1} {\{{\it k}-(0.5\pm\delta)\}^2+\kappa^2}
\end{displaymath}
\linebreak
\begin{figure}[hbt]
 \epsfxsize=3.0in
 \epsfbox{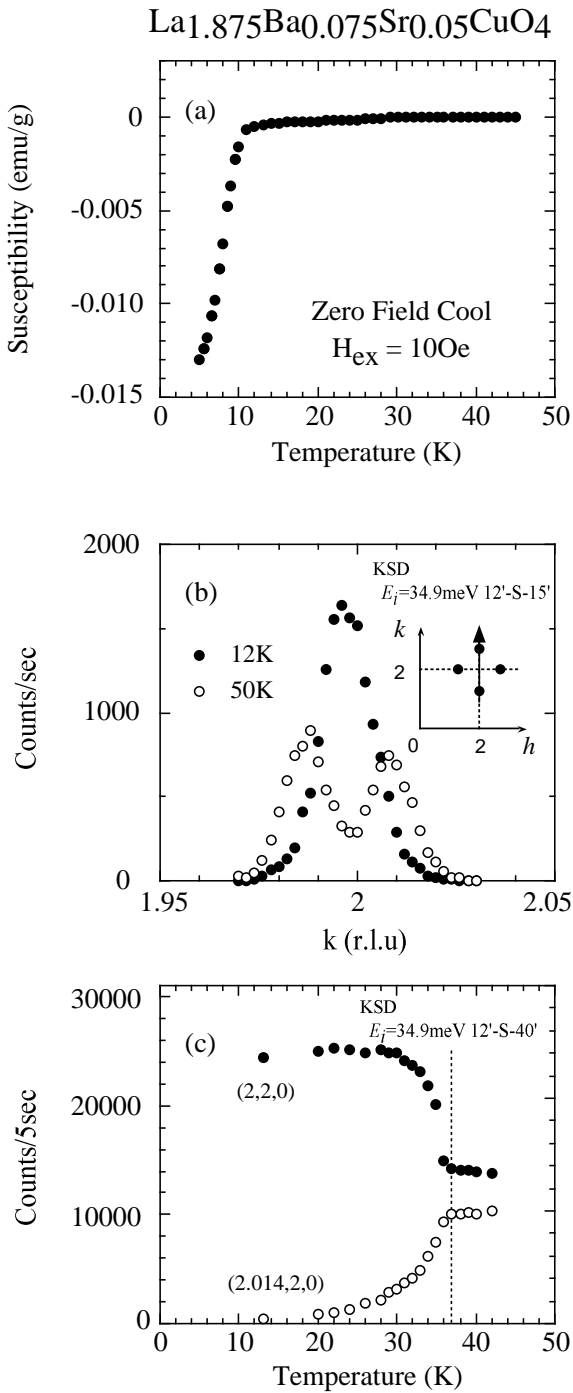}
  \caption{(a) Magnetic susceptibility measured in an applied field of 10 Oe after zero field cooling process. (b) Peak profiles along (2,{\it k},0) at 12K (closed circles), below $T_{d2}$, and 50K(open circles), above $T_{d2}$. The scan trajectory is shown in the inset. (c) Temperature dependence of the intensities at (2,2,0)(closed circles) and (2.014,2,0)(open circles).}
\end{figure}
\noindent
are fitted to the profile at 4.5 K, convoluting with the instrumental resolution. Where $\delta$ is the incommensurability corresponding to the half distance between the IC peaks and $\kappa$ is the peak width in half width at half maximum. 
Evaluated $\delta$ is 0.120$\pm$0.001, which is identical to that of the orthorhombic La$_{1.88}$Sr$_{0.12}$CuO$_{4}$~\cite{Kimura99}. The peak width is resolution limited, $\kappa$$\leq$0.0005$\AA$$^{-1}$. Furthermore, the peak intensity normalized by the sample volume is approximately six times larger than that in La$_{1.88}$Sr$_{0.12}$CuO$_{4}$ observed under identical experimental conditions.$^{11,12)}$

Figs.3 (a) and (b) show the elastic profiles 
\linebreak
\begin{figure}[hbt]
\epsfxsize=3.2in
\epsfbox{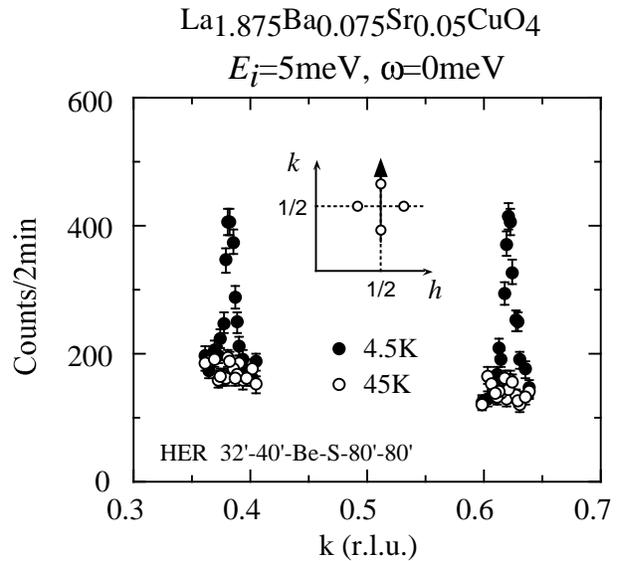}
\caption{Static IC peak profile along (0.5,{\it k},0) at 4.5(closed circles) and 45K(open circles) Inset shows the scan trajectory.}
\end{figure}
\noindent
for the paired IC peaks with the scan depicted in the inset of Fig.3(a). The center of each IC peak is found to locate at {\it h}=0.500(1) or 0.499(1), indicating the IC wave vector is nearly parallel/perpendicular to Cu-O bond direction. In fact, the deviation angle of IC wave vector from the Cu-O bond direction is negligible small (0.2$\pm$0.1$^{\circ}$) compared to that of $\simeq$3$^{\circ}$ in La$_{1.88}$Sr$_{0.12}$CuO$_{4}$$^{5,8)}$ and La$_2$CuO$_{4+y}$~\cite{YoungLee99} Same result is obtained for La$_{1.48}$Nd$_{0.4}$Sr$_{0.12}$CuO$_{4}$ with LTT phase.~\cite{Tranquada96} The peak positions reported for La$_{1.88}$Sr$_{0.12}$CuO$_{4}$ are shown by arrows in Figs.3 (a) and (b). Similar to the case of $k$-scan the peak widths along {\it h}-direction are also resolution limited. 

\section{Discussion}
Neutron scattering study on La$_{1.875}$Ba$_{0.075}$Sr$_{0.05}$CuO$_4$ shows the intense magnetic peaks at (0.5,0.5$\pm$$\delta$,0) with $\delta$=0.12. This result demonstrates that the magnetic order is commonly stabilized at the total hole concentration {\it p}$\approx$1/8 in hole-doped La-214 systems. The incommensurability $\delta$ is consistent with those of other systems with similar hole concentration. However, present result in LTT phase shows a clear contrast in the peak intensity as well as positions, compared with La$_{1.88}$Sr$_{0.12}$CuO$_4$ in LTO phase.

Assuming a common spin structure in LTO and LTT phases, we discuss the possible reasons for the difference in intensity. In the framework of stripe model, the elastic/quasi-elastic peak intensity is considered to be enhanced in LTT structure due to the pinning of the hole stripes along the Cu-O bond. In other words, the spectral weight of spin fluctuations at low-energy region shifts to the elastic side in LTT phase. Furthermore, the static stripe order pinned by LTT lattice potential is expected to be of long range and 3-dimensional. Indeed, the 3-dimensional correlation of charge order is observed in La$_{1.48}$Nd$_{0.4}$Sr$_{0.12}$CuO$_{4}$.~\cite{Zimmermann98} 
\linebreak
\begin{figure} [hbt]
\epsfxsize=3.0in
\epsfbox{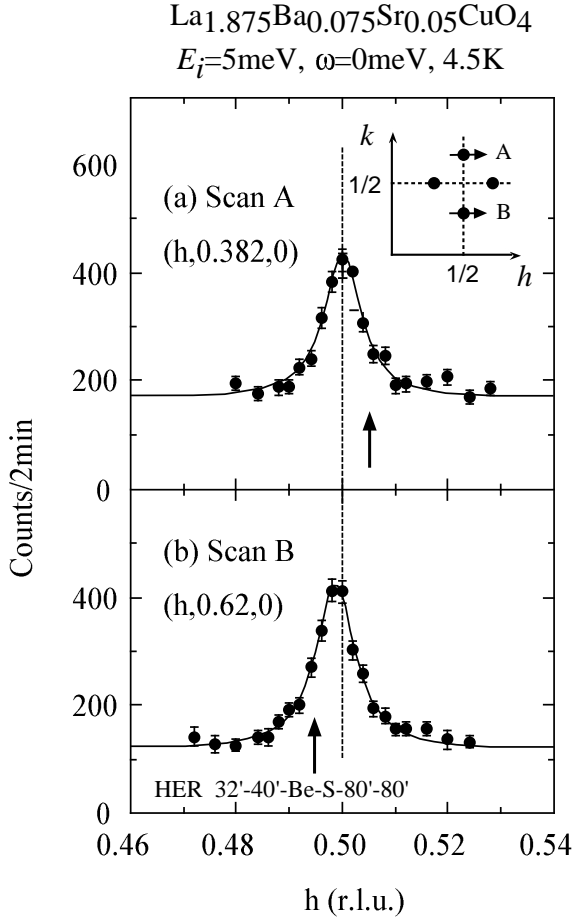}
 \caption{Elastic peak profiles along {\it h}-direction are shown in (a) and (b). In the figures, the peak positions observed in La$_{1.88}$Sr$_{0.12}$CuO$_{4}$ are indicated by arrows.~\cite{Kimura00} Scan trajectories are depicted in the inset of (a).}
\label{10K}
\end{figure}
\noindent
On the other hand, in the orthorhombic La$_{1.88}$Sr$_{0.12}$CuO$_4$ the correlation perpendicular to CuO$_2$ planes is rather short ranged.~\cite{Kimura00} Thus, in addition to the energy-shift, the peak broadening along {\it l}-direction may be one of the reasons for the weaker peak intensity of LTO phase compared with LTT. Note that recent high-resolution neutron powder diffraction measurements on La$_{1.875}$Ba$_{0.125-x}$Sr$_{x}$CuO$_4$ (LBSCO) system suggest the existence residual LTO phase in the sample for {\it x}$\leq$0.075 below {\it T}$_{d2}$.~\cite{Lappas00}. If this is the case, the difference in the magnetic peak intensity between LTT and LTO phases becomes larger than that determined in the present study.

Concerning to the peak positions, the square(rectangular) arrangement is confirmed in LTT(LTO) phase. This result suggests a close relationship between the peak position, the direction of wave vector and the crystal structure, the buckling pattern of Cu-O plane. Further experiments are needed to elucidate the relationship among the spin/charge order, the crystal 
\vspace{5cm}
structure and the superconductivity over a wide hole concentration. For this purpose, systematic study on LBSCO system is very helpful because the crystal structure can be changed by controlling the Ba/Sr ratio. Besides, this system is free from the rare-earth moment, so that the detailed investigation of low-energy spin dynamics as well as the out-of-plane correlation is possible. We have grown single crystals in this system at different doping level for the future neutron scattering experiments.

\section*{Acknowledgements}
We thank G.Shirane, H. Kimura and J.M.Tranquada for stimulating discussions. We acknowledge K.Nemoto and S.Watanabe for their technical support on the neutron scattering experiments at JAERI. This work was supported by a Grand-In-Aid for Scientific Research from Japanese Ministry of Education, Science, Sports and Culture.

\end{document}